\documentclass[preprint, 1pt]{elsarticle}
\journal{Annals of Physics}
\usepackage{bm}
\usepackage{latexsym}
\usepackage{amsfonts,amssymb,amsmath}
\usepackage{graphicx,epsfig}
\usepackage{psfrag}
\usepackage{subfigure}
\usepackage{color}
\usepackage{ulem}
\newcommand{\pauli}{{\sigma^{j}}}
\newcommand{\be}{\begin{equation}}
\newcommand{\ee}{\end{equation}}

\begin{document}
\begin{frontmatter}

\title{Floquet analysis of pulsed Dirac systems: A way to simulate rippled graphene}

\author{Tridev Mishra} 
\address{Department of Physics, Birla Institute of Technology and
  Science, Pilani 333031, India.\\ tridev.mishra@pilani.bits-pilani.ac.in}

\author{Tapomoy Guha  Sarkar}
\address{Department of Physics, Birla Institute of Technology and
  Science, Pilani 333031, India.\\ tapomoy1@gmail.com}  

\author{Jayendra  N. Bandyopadhyay}

\address{Department of Physics, Birla Institute of Technology and
  Science, Pilani 333031, India.\\jnbandyo@gmail.com}

\begin{abstract}

The low energy continuum limit of graphene is effectively known to be
modeled using Dirac equation in (2+1) dimensions. We consider the
possibility of using modulated high frequency periodic driving of a
two-dimension system (optical lattice) to simulate properties of rippled
graphene.  We suggest that the Dirac Hamiltonian in a curved
background space can also be effectively simulated by a suitable
driving scheme in optical lattice. The time dependent system yields, in the approximate limit of high frequency pulsing, an effective time independent
Hamiltonian that governs the time evolution, except for an initial and
a final kick. We use a specific form of $4$-phase pulsed forcing with
suitably tuned choice of modulating operators to mimic the effects of
curvature. The extent of curvature is found to be directly related to
$\omega^{-1}$ the time period of the driving field at the leading
order. We apply the method to engineer the effects of curved
background space. We find that the imprint of curvilinear geometry
modifies the electronic properties, such as LDOS, significantly. We
suggest that this method shall be useful in studying the response of
various properties of such systems to non-trivial geometry without
requiring any actual physical deformations.
\end{abstract}

\end{frontmatter}

%\pacs{37.10.Vz}{Mechanical effect of light on atoms, molecules, electrons, and ions} 
%\pacs{37.10.Jk} {atoms in optical lattice}
%\pacs{04.62.+v}{Quantum fields in curved spacetime}
%\pacs{73.22.Pr}{Electronic properties of Graphene}

\section{Introduction}
\label{sec:intro}

Quantum systems subjected to high-frequency periodic driving have
become a prominent feature of quantum simulation studies \cite{Bloch,
  Y.J.Lin}.  These studies are mostly aimed at modelling various
unique condensed-matter systems \cite{Greiner,Hadzi}.  Field induced
driving \cite{Dalibard.Gerbier}, or that generated through mechanical
straining, for instance, in graphene
\cite{VozKats,GuineKats,GonzalezGunie} have demonstrated their ability
to create novel gauge structures and modify the energy spectra.  Such
driving schemes have hence become increasingly popular in cold atom
and ion-trap systems as a means of implementing effective potentials
that could simulate magnetic fields or spin-orbit couplings
\cite{Sorensen,Lim,Hemmerich,Creffield,
  Bermudez,Struck,Hauke,Anderson}. The theoretical formalism
underlying these driven quantum systems relies on a time dependent
forcing that synthesizes an effective approximate time
independent Hamiltonian
\cite{NautsWyatt,Maricq,Shirley,Sambe,Grozdanov,Avan,Rahav,JNB-TGS}. A recent
trend in these investigations has been inclined towards looking at a
variety of driving schemes to explore potentially interesting
Hamiltonians \cite{Dalibard}.

  In much of the last decade two areas have witnessed rapid progress,
  namely, the physics of graphene with its
  applications \cite{CastroNeto} and ultra cold atoms in optical
  lattices \cite{ImBloch}.  Interest in the former is driven by the
  realization of a perfectly flat two-dimensional (2D) system and the unique
  physics observed in the material due to its relativistic dispersion
  relation \cite{Wilson,Neto,Zhou}. Optical lattices, on the other hand, has offered an
  indispensable simulator for realizing many-body condensed matter
  phenomena and noting their response to a highly controllable
  variation of system parameters.  This has motivated a significant
  advancement in the efforts to simulate graphene like systems in optical
  lattice \cite{ZhuWang,Englert,Hou,Juzeli,Wunsch,Shen,Asano,Tarruell,Polini,Jotzu}.

 Graphene is noted to show exotic properties, either under mechanical
 strain, curvature or possessing defects such as dislocations
 \cite{VozFuller,VozKats,GuineKats,GonzalezGunie,VozJuan,IorioWeyl,IorioUnruh,VozCorti,Vozmidgap}.
 These studies use a continuum model of Dirac fermions in curved
 (2+1) dimensions in the limit of low energy excitations.  This has
 also found possibilities of experimental investigation in the cold
 atom/optical lattice setup with the objective of studying
 relativistic electrodynamics in the presence of gravity
 \cite{Lewenstein}. 
 
 The experimental realization of such systems has
 presented technical difficulties arising from the spin-like and
 position dependent nature of the nearest-neighbor hopping amplitude
 in their Fermi-Hubbard Hamiltonian. The essential requirement is the
 coupling of an artificial non-abelian gauge field to the ultra cold
 fermionic atoms in the optical lattice (near half-filling) giving
 rise to the appropriate effective dynamics
 \cite{Satija,Zoller,Ohberg,Gunter,Goldman,Ruseckas,Bermude,Rev,Panahi}.

 A key ingredient of all such simulations involves the generation of
 artificial gauge fields in optical lattices through periodic driving
 or `shaking' \cite{Struck,Kengstock}. We propose the use of a certain
 driving scheme to obtain an effective curved graphene model in the
 optical lattice setup.  The key distinction of our proposed scheme
 from similar works \cite{Lewenstein} is our use of pulse
 sequences with suitably chosen modulating operators, as described in
 \cite{Dalibard}, to generate the effects of smooth driving. This is
 suggested as an alternate scheme to circumvent difficulties arising
 from the complicated form of the effective tunneling parameter in
 conventional treatments.  An added advantage of this method is the
 easy correspondence afforded by it between the continuum and the
 lattice using a suitable map relating the operators in the two
 pictures.
 
In this work, we outline a scheme for the generation of an
approximate effective Hamiltonian using periodic time dependent forcing on an hexagonal optical lattice
having the Bloch band topology of flat graphene such that the resulting static effective system mimics the
features of a curved background. To compare with  the Dirac equation in
curved (2+1) dimensional background  we consider a  metric with a conformally
flat spatial part. The Dirac equation in this curved background is
cast in a Hamiltonian form to allow easy comparison with the lattice
Hamiltonian in the continuum limit. The effect of the background
curvature is noted.  We have found an effective approximate time-independent 
Hamiltonian which is obtained from a specific high frequency time-periodic driving of the flat space Dirac Hamiltonian. This effective Hamiltonian is found to be identical to the Dirac Hamiltonian in curved space at the leading order.

We also note the direct correspondence between the nature and the periodicity of the driving to the form and the extent of curvature. The modification of the electronic properties, specifically Local Density of States (LDOS) is studied in low energy regimes near the Fermi points.

\section{Formalism}
\label{sec:formalism}
 
 \subsection{Massless Dirac Equation in curved (2+1) D space}
 
 We consider the effect of the curvature of the background space on the massless Dirac equation. The curved space Dirac Hamiltonian is believed to govern the 
 quasi-particle (i.e., the massless Dirac fermion)  dynamics in the continuum limit of the low energy approximation for graphene sheets with curvature. In the subsequent sections, we shall elaborate upon our intent to replicate such systems in the framework of optical lattice simulation.
 
The Dirac equation in (2+1) dimensional space-time has been studied in various contexts and has a well defined formalism
\cite{Pollock,Sucu,Khalilov,Gitman,DasGupta,Proy}. This section provides a brief overview of this as relevant to
our work. We consider a  (2+1) dimensional space-time as the backdrop for our analysis.
We choose a space-time metric of the form
\begin{eqnarray}
\label{metric}
 ds^{2}=  dt^2 -e^{-2\Lambda(x,y)}(dx^2 + dy^2).
\end{eqnarray}
where $t$ represents the time coordinate, $x$ and $y$ are the spatial
isothermal cartesian coordinates, and $ e^{-2\Lambda(x,y)}$ denotes the 
conformal factor.  
We note here that the two-dimensional spatial part of this metric $diag\, (1 , -e^{-2\Lambda(x,y)}, -e^{-2\Lambda(x,y)} )$ is completely general in representing two-dimensional curved surfaces. This metric has been used in the context of
studying Dirac equation coupled to curved space-time
\cite{BrillWheeler.RevMP23} with a distribution of defects, for
instance, in the case of corrugated graphene sheets \cite{VozKats}.

The Dirac equation in curved space-time takes the form
\begin{eqnarray}
\label{diracm0}
 i\gamma^{\mu}(x)(\partial_{\mu} + \Gamma_{\mu}(x))\psi=0
\end{eqnarray}
The spin connection term, $\Gamma_\mu(x)$, is given by  \cite{BrillWheeler.RevMP23}
\begin{equation}
 \Gamma_\mu(x) = g_{\lambda\alpha}(e^i_{\nu,\mu}E^\alpha_i - \Gamma^\alpha_{\nu\mu})s^{\lambda\nu} + a_\mu\mathcal{I}
\end{equation}
where ${{e^i_{\nu}}}$ and ${{E^\alpha_i}}$ denote the usual vielbeins
and their inverses respectively, $\Gamma^\alpha_{\nu\mu}$ are the
Christoffel connection coefficients and $s^{\lambda\nu}$ are the
generators of spinor transformation in curved space-time. This expression illustrates the indeterminacy of the connection term to
upto a constant $a_{\mu}$. Hence $\Gamma_\mu$ has an arbitrary trace
\cite{BrillWheeler.RevMP23}. This offers a gauge freedom which can be
exploited depending on the nature of the problem. We take the standard
choice for $\Gamma_{\mu}$ as
\begin{equation}
 \Gamma_\mu(x) = \frac{1}{2}g_{\lambda\alpha}(e^i_{\nu,\mu}E^\alpha_i - \Gamma^\alpha_{\nu\mu})s^{\lambda\nu}
\label{fgamma}
\end{equation}
with,
\begin{equation}
 s^{\lambda\nu}(x)= \frac{1}{2}[\gamma^\lambda(x),\gamma^\nu(x)].
\end{equation}
The $\gamma$ matrices with  space-time indices are related to the usual Dirac matrices in flat space
by $ \gamma^{\mu}(x) = E^{\mu}_{i} (x)\gamma^{i}$. We choose the following representation using the Pauli matrices for  the
$\gamma^i$s
\begin{equation}
 \gamma^0 = \sigma^z~~~~\gamma^1 = i\sigma^y~~~~\gamma^2 = -i\sigma^x.
\end{equation}
In our choice of representation, $\sigma^z$ is diagonal and $\sigma^y$ is complex.

The spin connection components, for our metric [see Eq.\eqref{metric}], are given as
\begin{equation}
 \Gamma_1(x)= \frac{i}{2}\frac{\partial\Lambda(x,y)}{\partial y}\sigma^z,~\Gamma_2(x)= -\frac{i}{2}\frac{\partial\Lambda(x,y)}{\partial x}\sigma^z.
\end{equation}
The massless Dirac equation in curved (2+1) space-time, can hence be written as
\begin{equation}
\begin{split}
&\biggl[i\sigma^z\frac{\partial}{\partial t}-
    e^{\Lambda(x,y)}\biggl(\sigma^y\frac{\partial}{\partial
      x}-\sigma^x\frac{\partial}{\partial y}\biggr)\\ &+
    \frac{e^{\Lambda(x,y)}}{2}\biggl(\frac{\partial\Lambda (x,y)}{\partial
      y}\sigma^x - \frac{\partial\Lambda(x,y)}{\partial
      x}\sigma^y\biggr)\biggr]\psi=0,
\end{split}
\end{equation}
where we have  used eqns. \eqref{diracm0} and
\eqref{fgamma}. This equation can be recast in an  
explicitly Hamiltonian form by breaking the manifestly covariant form as
\begin{equation}
\label{DiracHam}
i\frac{\partial\psi}{\partial t}=e^{\Lambda(x,y)}
\biggl[-i{{\sigma^{j}}}\partial_{j} -
  \frac{i}{2}\biggl(\frac{\partial\Lambda(x,y)}{\partial
    y}\sigma^y + \frac{\partial\Lambda(x,y)}{\partial x}\sigma^x
  \biggr)\biggr]\psi. 
\end{equation}
The entire operator acting on $\psi$ in the RHS of 
the above equation may be interpreted as the Dirac Hamiltonian in curved space.
This Hamiltonian is required to be synthesized using
the driven optical lattice. As will be shown later it 
is possible to formulate a driving 
scheme which does exactly this. In the following section
we discuss the procedure for obtaining an effective time-independent
Hamiltonian for a periodically driven systems. This shall find appropriate
implementation in optical lattices.
 
\subsection{Periodic Pulsing and Effective Hamiltonians}
 
In the study of quantum systems having periodic time dependent
Hamiltonians \cite{Shirley,Sambe},  a special category is devoted to the
class of systems where the system is subjected to high frequency periodic forcing \cite{NautsWyatt}. The theoretical treatment of such systems has
its roots in the study of similar classical
systems \cite{Landau,Percival}.  The literature suggests various routes to arrive at  an
effective time-independent Hamiltonian
\cite{Grozdanov,Rahav,Maricq,Avan}. The traditional practice of using Cambell-Baker-Hausdorff (CBH) expansion or Trotter expansion to study Floquet systems have certain inherent defects \cite{Rahav,JNB-TGS}. 
A recent approach \cite{Dalibard}, inspired by \cite{Rahav}, forms the basis of our formalism. It uses the
idea of engineering effective Hamiltonians by applying carefully
selected periodic driving schemes to quantum systems, geared towards
generating desired effective static systems.

The technique may be outlined as follows. One considers a
time-periodic Hamiltonian $H(t)$ that can be written  as
\begin{equation}
 H(t) = H_0 + V(t),
\end{equation}
where $H_0$ is time independent and
$V(t)$ is the periodic time dependent part such that $V(t + T) =
V(t)$. The idea is to decompose the unitary time-evolution operator $U(t_{i},t_{f})$, which governs the dynamical evolution of the system between time slices $t_i$ and $t_f$ in the following form
\begin{equation}
\label{evoperator}
U(t_{i},t_{f}) = e^{-i {F}(t_{f})}e^{-i {H}_{eff}(t_{f}-t_{i})}e^{i {F}(t_{i})}. 
\end{equation}
Here ${H}_{eff}$ is a time-independent effective Hamiltonian and
${F}(t)$ is a time dependent Hermitian operator with
${F}(t+T)={F}(t)$.  It is important to note here that
${H}_{eff}$ is independent of both $t_i$ and $t_f$, which have
been transferred into the ``Kick'' terms $e^{i{F}(t_{i})}$ and
$e^{-i{F}(t_{f})}$ respectively.  It is generally not possible to
extract the operators $F(t)$ and $H_{eff}$ in closed analytic form
except for some special cases. However, if  the driving frequency $
\omega = 2\pi/ T$ is high, one can consider a perturbative expansion
using the small parameter $1/\omega$. The  expansions for ${H}_{eff}$ and ${F}$ are 
\begin{equation}
{H}_{eff}=\displaystyle\sum_{0\leq n<
   \infty}\frac{1}{\omega^{n}}{H}^{(n)}~~~{F}=
 \displaystyle\sum_{1\leq n< \infty}\frac{1}{\omega^{n}} {F}^{(n)}.
\label{eq:pert}
\end{equation}
The time evolution equation for $U(t_i, t_f)$ given by $i \partial_t U(t)  = H U(t) $  yields
\begin{eqnarray}
 {H}_{eff}= e^{i{F}(t)}{H}e^{-i{F}(t)} + i\frac{\partial}{\partial t}\left(e^{i{F}(t)}\right)e^{-i{F}(t)}.
\end{eqnarray}
This may be expanded as a perturbation series in $ 1/\omega$ using
Eq.\eqref{eq:pert}. We note that the
operators $F^{(n)}$ are all periodic with zero mean so that we have 
\be 
\langle F^{(n)} \rangle = 0, ~~~~~~F^{(n)}(t+T) = F^{(n)}(t).  
\ee

The periodic time-dependent operator $V(t)$ can be expanded in a
Fourier series as 
\begin{equation}
 {V}(t)= {V}_0 + \displaystyle\sum_{1\leq n<\infty}\hat{V}_{n}e^{in\omega t} + \displaystyle\sum_{1\leq n<\infty}\hat{V}_{-n}e^{-in\omega t}.
\end{equation}
At a given order of perturbation in
$(1/\omega)$ in Eq. (\ref{eq:pert}), one retains the time independent average in $H_{eff}$
and adjusts $F(t)$ to annihilate any time dependence. The procedure is
repeated at each order and results obtained at the previous order is
incorporated in the subsequent orders.  This allows us to determine
$H^{(n)}$ and $F^{(n)}$ up to the desired accuracy. This yields the following expression for the effective Hamiltonian \cite{Dalibard}
\begin{equation}
\begin{split}
\label{effective}
 {H}_{eff} &= {H}_0 + {V}_0+ \frac{1}{\omega}\displaystyle\sum_{n=1}^{\infty}\frac{1}{n}[\hat{V}_n,\hat{V}_{-n}] \\
 &+ \frac{1}{2\omega^2}
 \displaystyle\sum_{n=1}^{\infty}\frac{1}{n^2}\biggl(\biggl[[\hat{V}_n,H_0],\hat{V}_{-n}\biggr]+h.c. \biggr) + \mathcal{O}(\omega^{-3}).
\end{split}
\end{equation}

We shall now focus on a specific kind of forcing potential. 
The driving potential $V(t)$ shall be considered to be a sequence of pulses that
repeat periodically. The choice of the number of phases in a given pulse sequence dictates the form of the effective Hamiltonian. This offers a wide variety of possibilities up to a given order $\omega^{-1}$ in the perturbation expansion.

Let us consider a general $N$-phase pulse sequence, with period $T$, of the form 
\be 
V(t) =
\displaystyle\sum_{r=1}^{N}f_r(t) V_r
\label{pulseQ}
\ee 
where $f_r$ denotes a square profile such that 
\be
f_r(t) = \begin{cases} 1, & (r-1)T/N \leq t \leq rT/N,\\
0, & {\rm elsewhere.}
\end{cases}
\ee
Here, $V_r$ are 
arbitrary operators that are free to be chosen as per ones requirement. Each
phase lasts for a duration of $T/N$. We also impose the condition $\sum\limits_{r=1}^N V_r = 0$.

The time-dependent Hamiltonian for such a choice of driving is then
\be 
H(t)= H_0 + \displaystyle\sum_{r=1}^{N}f_r(t) V_r . 
\ee 
Using the Fourier series expansion this can be written as 
\be 
H(t) = H_0 + \displaystyle\sum_{n\neq 0}\hat{V}_{n}e^{in\omega t}, 
\ee 
where
\be
\hat{V}_n = \frac{1}{2\pi
  i}\displaystyle\sum_{r=1}^{N}\frac{1}{n}e^{- 2\pi i
  nr/N}(e^{2\pi i n/N}-1) V_r.  
\ee 
It is possible to use Eq.\eqref{effective} at this stage to obtain a generic expression for
the time-independent effective Hamiltonian for the kind of driving
given in Eq. \eqref{pulseQ} (refer Eq. (30) in \cite{Dalibard}).

Given the flexibility of choosing the number of phases and also the
modulating operators, a wide variety of effective Hamiltonians can be
generated. The next section deals with one such choice that
enables us to design the required gauge field to simulate the physics
of curved graphene in optical lattices. The usefulness of such a pulsing scheme is demonstrated by showing its equivalence to an optical
lattice shaken/modulated by a smooth driving. 

The modulation scheme used in standard optical lattices does not consist of such pulsing and instead uses smooth driving. The effective Hamiltonian obtained for smoothly modulated optical lattices carries an imprint of the modulation frequency through the renormalized hopping term (which is a function of $\omega$). On Taylor expanding the hopping parameter as a series in $\omega^{-1}$, 
this effective Hamiltonian matches with the one obtained by a pulsing scheme at the leading order \cite{Dalibard}.

\subsection{Simulating graphene in curved space: Optical Lattice Scheme}

As mentioned previously, the use of hexagonal optical lattices
to simulate Dirac cones and massless Dirac fermions is well 
established. In such a system the application of a time-dependent
sinusoidal modulation can be used to obtain novel gauge effects 
in an artificial time-averaged manner. The possibility of doing this 
using the method discussed in the previous section is elaborated here.

Among the wide range of choices that do exist, our problem lends
itself rather neatly to a $4$-phase pulse sequence with modulation of
the Hamiltonian given by
\begin{equation}
\label{4phase}
 \mathcal P_4: \{{H}_0 + {A}, {H}_0 + {B}, {H}_0 - {A}, {H}_0 - {B}\}  
\end{equation}
This compares to Eq.\eqref{pulseQ} for $N=4$ with ${V}_1 =-{V}_3= {A}$
and ${V}_2=-{V}_4={B}$, where $A$ and $B$ are suitable operators.
As discussed in the last section, this is equivalent to a smooth driving  of the form 
\be 
V(t)= {A} \cos(\omega t) + {B} \sin(\omega t). 
\ee  
This choice of the time-dependent
potential yields the following effective Hamiltonian \cite{Dalibard}
\begin{equation}
\label{4phaseH}
\begin{split}
  {H}_{eff} &= {H}_0 + \frac{i}{2\omega}[{A}, {B}] + \\
 &\frac{1}{4\omega^2} \left( [[{A},{H}_0], {A}]  
 +  [[{B},{H}_0],{B} ]\right)+\mathcal{O}(1/\omega^3)
 \end{split}
\end{equation}
It is significant in our context to note that the expression for $ {H}_{eff}$
has both first order and second order terms in $\omega$ with the
appropriate commutator brackets. The freedom in the choice of ${A}$
and ${B}$ allows us to engineer the desired effective Hamiltonian.

The periodic driving scheme has a small parameter $\omega^{-1}$, the
time-period of forcing. It is our contention that it
is possible to use the formalism of generating effective approximate
Hamiltonians, through a choice of suitable operators $A$ and $B$ as
mentioned in Eq.\eqref{4phase}, to reproduce a Dirac Hamiltonian 
in  curved space. This would involve choosing an appropriate pulsing scheme. 

We note that the low energy limit of a continuum approximation of graphene, 
as simulated in the lattice, has the Hamiltonian of the form \cite{CastroNeto} 
\be
H_{\rm G} = -iv_F \pauli  \partial_{j}
\label{graphene}
\ee in units of $\hbar$, where $v_F$ is the Fermi velocity and $\partial_{j} = (\partial_x,\partial_y)$ is the gradient operator in 2-dimensions.
We shall subsequently work in
units where $v_F = 1$. This motivates us to consider the primary Hamiltonian 
in our analysis as $-i\pauli  \partial_{j}$. The discussion here solely
employs the continuum formalism for the operators and the mapping to the second quantized forms for the operators and the Hamiltonians are only introduced later in the final section. 

Let us consider a driving scheme with $H_0 = -i\pauli \partial_{j}$,
the Dirac Hamiltonian in flat space and choose the operators ${A}$ and
${B}$ of the form
\begin{equation}
 {A} = \pauli  {{\alpha}_{j}}~~~~~~~~ {B} = \sigma^{k}  {{\beta}_{k}}
 \label{operators}
\end{equation}
where, ${{\alpha}_{j}} = [i\partial_y,-i\partial_x,0]$ and
${{\beta}_{k}} = [0,0,-f(x, y )]$.  With this choice,  Eq.\eqref{4phaseH} yields  an approximate effective Hamiltonian $H_{eff}$  up to order $\omega^{-1}$ given by
\begin{equation}
\label{eq:effectiveh}
\begin{split}
H_{eff}  &= \frac{1}{2}\left[-i\left(1+\frac{f(x,y)}{\omega}\right)\pauli  \partial_{j}\right] \\ &-\frac{1}{2}\left[i\pauli  \partial_{j}\left(1+\frac{f(x,y)}{\omega}\right)\right]
\end{split}
\end{equation}  
 For large $\omega$ this is a good approximation. 
 The term of $\mathcal{O} (\omega^{-2})$ is significantly
suppressed and manifests as non-trivial spin-orbit couplings and maybe
ignored for our present analysis. With a substitution
$$ e^{\Lambda(x,y)} = \left(1+ \frac{f(x,y)}{\omega}\right)$$
we have 
\be
\label{eq:effectiveh1}
H_{eff} = \frac{1}{2}[ -ie^{\Lambda(x,y)}\pauli  \partial_{j}  - i\pauli  \partial_{j}~e^{\Lambda(x,y)}]
\ee
such that the entire expression is in terms of $\Lambda(x,y)$ instead of 
$f(x,y)$.
%it can be readily shown that one recovers the curved space Dirac Hamiltonian 
%in \eqref{DiracHam}.
The Hamiltonian in Eq.\eqref{eq:effectiveh1} can be further simplified and explicitly written as follows
\be
H_{eff} =e^{\Lambda(x,y)}\biggl[-i\pauli  \partial_{j} -\frac{i}{2}\biggl(\frac{\partial\Lambda(x,y)}{\partial
    y}\sigma^y + \frac{\partial\Lambda(x,y)}{\partial x}\sigma^x
  \biggr)\biggr]
\label{eq:effective}
 \ee
We seek to map this effective time-independent Hamiltonian that is obtained from the original time-dependent Hamiltonian to the Dirac Hamiltonian in curved space. The function $\Lambda(x,y)$ appearing here is expected to be mapped to the metric in some fashion in the equivalent curved space description. 

Comparing Eq.\eqref{eq:effective} and  Eq.\eqref{DiracHam} we establish the
correspondence between the periodically driven effective system and 
a curved space description. The function $\Lambda (x, y) $ that depends on  the periodic  driving scheme is now seen  to appear in  the conformal factor of the metric in the curved space picture. The quantity of geometrical interest describing 2D curved surface is the Gauss curvature $K(x,y)$ given by 
\be
\label{Gauss}
K(x,y) =   e^{2\Lambda} \nabla^2 (\Lambda)
\ee
This scalar function has complete information about the curved 2-D surface.
Since $ \Lambda = \ln \left( 1 + \frac{f(x,y)}{\omega} \right)$ depends on the driving scheme $f$ and driving frequency $\omega$, the curvature shall depend on these directly. It is hence possible to reproduce the effects of curvature ($ K \neq 0$) by suitably manipulating the driving scheme. This completes the mapping between the two equivalent pictures.

In order to confirm that our model suitably mimics the properties
of curved graphene, it is required that some physical quantity
associated with it be computed and obtained experimentally. We regard the Local density of 
states (LDOS) to be a suitable candidate. In the following
we briefly recapitulate its significance and prescribe a method
for determining it theoretically.

The LDOS is a quantity of interest in the
study of electronic and transport properties of various condensed
matter systems. It offers information regarding the spatial variation
in the density of states over a region, arising out of local
disturbances, that can be verified experimentally using scanning
tunneling microscopy (STM) techniques.  It is therefore a physically
relevant parameter for our study.
Our analysis suggests that the electronic properties for a  periodically driven graphene
like optical lattice system, describable by a Dirac Hamiltonian, shall be 
the same as one expects for the same system in a curved background without any periodic 
forcing. 
  To compute the
LDOS \cite{adam}   one first needs to calculate the Green's function  for the system,
for the case of non-interacting electrons, as follows.  \be
\label{GREEN}
G(z,{\bf {r}},{\bf {r'}}) = \displaystyle\sum_{n}\frac{\psi_n({\bf
    {r}})\psi^*_n({\bf {r'}})}{(z -E_n)} \ee where, $z$ denotes a
complex energy variable, $\psi_n$ are energy eigenstates in coordinate
representation, $E_n$ represents the energy eigenspectrum and the sum
ranges over the $n$ eigenvalues of energy.  The expression for the
LDOS is given as 
\be \rho(\epsilon,{\bf {r}}) = -\frac{1}{\pi}Im
\displaystyle\sum_{n}\frac{|\psi({\bf{r}})|^2}{(\epsilon +i\delta
  -E_n)} \ee which may be written as  \be
\label{LDOS}
{\rm LDOS} = \rho(\epsilon,{\bf {r}})= -\frac{1}{\pi}Im [G(\epsilon +i\delta,{\bf {r}},{\bf {r'}})]
\ee
We shall compute the LDOS numerically using the spectrum of the Hamiltonian in Eq. \eqref{eq:effective} and compare it with the flat space case where $\Lambda = 0$.

\section{Results and Discussion}
 
 The study of alterations to the electronic properties of graphene
 sheets as a result of deformation, curvature, defects or impurities
 focusses chiefly on the modifications to the LDOS or the appearance of a
 gap at the Fermi points \cite{VozJuan,GuineKats, Zhou,
   Gui,Han,Snyman,Tiwari, Coletti,Gomes}. These works discuss the
 possibility of opening a band gap in graphene at the Dirac point,
 which is known to be topologically protected by inversion and time
 reversal symmetries \cite{Haldane,Herbut, Zhu}. The presence of
 perturbations that respects these discrete symmetries can only move
 the Fermi points but not create a gap \cite{manes}. A hybridization
 of the Fermi points with opposite topological charge (winding number)
 allows a subsequent opening of gap \cite{Montambaux}.

In our present analysis we attempt to examine the effect on the LDOS
for graphene-like optical lattice under a periodic driving. The
approach has similar motivations to earlier studies on LDOS in rippled
graphene \cite{VozJuan}.  The principal difference being that our
system does not involve taking a graphene sheet with any curvature or
defects but imparting curved-graphene properties to an optical
lattice via pulsing. The choice of the driving scheme function $f(x,y)$ that 
maps to the conformal factor in the metric is taken as 
\be
\label{function}
f(x,y)= x^2 + y^2. 
\ee 
This choice of the driving scheme is used to
compute the curvature according to Eq. \eqref{Gauss} and yields a constant Gaussian curvature $K(x,y) = \frac{4}{\omega}$. Thus the curvature turns out to
be inversely proportional to the driving frequency $\omega$. Hence, with
our high frequency driving scheme (high frequency is a necessary
condition required for the convergence of the perturbation series in
Eq. \eqref{eq:pert}) we are able to model a small positive constant
curvature.

It is possible to write down the operators $A$ and $B$ of the
driving in the conventional second quantized notation. To do so we adopt a convention in which the hexagonal optical lattice Hamiltonian reads 
\be 
H_0 = J \displaystyle\sum_{\langle k,j \rangle}
\Psi^{\dagger}_{k+1,j}\sigma^x\Psi_{k,j} +
\Psi^{\dagger}_{k,j+1}\sigma^y\Psi_{k,j} - h.c. + H_{\rm on-site} 
\ee
where, $J$ is the plain hopping parameter, $a$ the lattice spacing,
  $ \Psi^{\dagger}_{k,j} =
  (\hat{a}^{\dagger}_{k,j},\hat{b}^{\dagger}_{k,j})$ creates a
  particle at the site $ (ka,ja)$ in some spin state. The operators $
  \hat{a}_{k,j}$ and $\hat{b}_{k,j}$ stand for the two triangular
  sub-lattices of the hexagonal lattice. The operators $A$ and $B$ in
  this convention, for the choice of $f(x,y)$ in Eq. \eqref{function},
  becomes 
\be 
\begin{split}
A &= -\frac{i}{2a}\displaystyle\sum_{\langle k,j \rangle}
  \Psi^{\dagger}_{k,j+1}\sigma^x\Psi_{k,j}
  -\Psi^{\dagger}_{k+1,j}\sigma^y\Psi_{k,j}-h.c.\\  
B &= -\sigma^z\frac{a^2}{2}\displaystyle\sum_{\langle k,j \rangle}
  k^2\Psi^{\dagger}_{k+1,j}\sigma^x\Psi_{k,j} +
  j^2\Psi^{\dagger}_{k,j+1}\sigma^y\Psi_{k,j} + h.c.
\end{split}
\ee
In the above expressions, we make use of the following map between continuum operators and those on the lattice as 
\begin{equation} 
\begin{split}
&-i\sigma^x\partial_x \equiv \frac{i}{2a}\displaystyle\sum_{\langle k,j \rangle} \Psi^{\dagger}_{k+1,j}\sigma^y\Psi_{k,j}-h.c. \\
&-i\sigma^y\partial_y \equiv \frac{i}{2a}\displaystyle\sum_{\langle k,j \rangle}\Psi^{\dagger}_{k,j+1}\sigma^x\Psi_{k,j}-h.c.
\end{split}
\end{equation}
and
\be 
x^2+y^2 \equiv \frac{a^2}{2}\displaystyle\sum_{\langle k,j \rangle} k^2\Psi^{\dagger}_{k+1,j}\sigma^x\Psi_{k,j} + j^2\Psi^{\dagger}_{k,j+1}\sigma^y\Psi_{k,j} + h.c.
\ee
The mapping between the continuum operators and their lattice counterparts enables the actual possibility of simulation of the Hamiltonian on the lattice.

\begin{figure}
\includegraphics[height=6.25cm, width= 7cm]{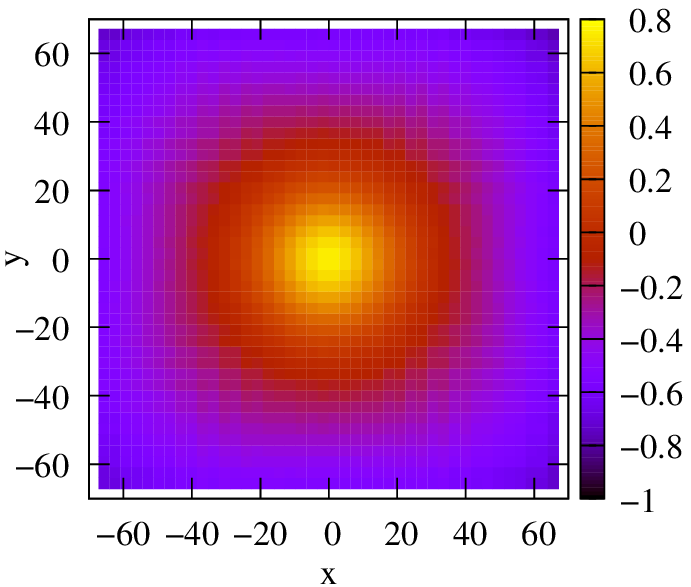}
\caption{(Color Online) Correction to the LDOS given by $\frac{\rho}{\rho_o}-1$ with $\rho$ being the LDOS for pulsed graphene and $\rho_o$ that for ordinary graphene.}
\label{fig:fig1}
\end{figure}

We investigate the nature of the LDOS for the Hamiltonian in Eq. \eqref{eq:effective} and look for the imprint of spatial curvature in its behavior.
The LDOS computations are performed for the choice of the driving scheme given in Eq. \eqref{function}. The expressions in Eq. \eqref{GREEN} and Eq. \eqref{LDOS} are evaluated numerically to estimate the LDOS. The Fig. \ref{fig:fig1} 
shows the modification to the LDOS for our system over that of normal graphene in flat space. The figure shows the quantity $(\rho/\rho_0) - 1$ plotted in the color contour map against the spatial coordinates $x$ and $y$. As seen in the figure, a large positive correction is centered at the reference origin indicating maximum increase in the number of available states per unit energy. This is a clear indication that electronic properties are significantly altered in our system. An $80\%$ correction is observed at the maxima for our choice of driving frequency which yields a $\omega^{-1}$ of $\sim 0.01$. We note that a similar behavior of the LDOS has also been observed in the study of graphene in curved space with positive curvature \cite{VozCorti}.

\section{Conclusion}

We conclude by noting that the use of periodic forcing to generate the effects of curved space on 2D quantum systems has a far reaching influence in theoretical studies and technological applications. The traditional Floquet analysis of periodically driven systems uses the CBH/Trotter expansion to find the effective static Hamiltonians. We use an alternative perturbative formulation using a pulsed driving scheme and found an effective approximate Hamiltonian. We show that the driving scheme can be chosen to simulate desired geometric properties on space. Our work particularly studies an optical lattice analogue for graphene in curved space. The massless Dirac equation and Hamiltonian in curved space, that model electronic behavior in curved graphene, are derived for a conformal metric. The same is shown to be obtained in a periodically driven hexagonal optical lattice having chosen the appropriate modulating operators. We go on to analyze the geometrical and physical features of the system, namely, the Gauss 
curvature and LDOS. These are computed for a particular choice of metric and deviations from the unperturbed system are noted. This opens up the possibility of synthesis of new systems in quantum simulators and the study of their physical properties.

\end{document}